\documentclass{article}
\usepackage{ijcai19}

\usepackage{siunitx} 

\usepackage{times}
\usepackage{soul}
\usepackage{url}
\usepackage[hidelinks]{hyperref}
\usepackage[utf8]{inputenc}
\usepackage[small]{caption}
\usepackage{graphicx}
\usepackage{amsmath}
\usepackage{amssymb}
\usepackage{booktabs}
\usepackage{algorithm}
\usepackage{algorithmic}
\newcommand{\mC}{\ensuremath{\mathcal{C}}}

\newcommand{\mI}{\ensuremath{\mathcal{I}}}

\newcommand{\be}{\begin{enumerate}}
\newcommand{\ee}{\end{enumerate}}
\newcommand{\bi}{\begin{itemize}}
\newcommand{\ei}{\end{itemize}}

\title{A mixed-integer linear programming approach for soft graph clustering}

\author{
Vicky Mak-Hau
\and
John Yearwood
\affiliations
School of Information Technology, Deakin University, Waurn Ponds, Vic. 3215, Geelong, Australia
\emails
\{vicky.mak, \ john.yearwood\}@deakin.edu.au
}

\begin{document}

\maketitle

\begin{abstract}
This paper proposes a Mixed-Integer Linear Programming approach for the Soft Graph Clustering Problem. This is the first method that simultaneously allocates membership proportion for vertices that lie in multiple clusters, and that enforces an equal balance of the cluster memberships. Compared to (\cite{Palla2005}, \cite{Derenyi2005}, 
\cite{Adamcsek2006}), the clusters found in our method are not limited to $k$-clique neighbourhoods. Compared to (\cite{Hope2013}), our method can produce non-trivial clusters even for a connected unweighted graph.

\end{abstract}

\section{Introduction}
\label{intro}
 
Consider an undirected graph $G = (V,E)$ with $V$ the set of vertices and $E = \{ (i,j) \ | \ i, j \in V, i < j\}$ the set of edges. Each edge $e \in E$ is associated with a weight $w_e \in \mathbb{R}$ that indicates the similarity between its two end vertices--the lager the weight, the more ``similar'' the two vertices are. The hard graph clustering (HGC) problem is to create distinct partitions (clusters, or, communities) of the set of vertices according to their similarities, i.e., to form $V_1,\ldots,V_k$, where $\bigcup_{i = 1,\ldots,k} V_i = V$, and $V_i \cap V_j = \emptyset$ for all $i,j \in \{1,\ldots,k\}$, $i \neq j$. For a thorough literature review of the graph clustering problems, see, e.g., \cite{Schaeffer2007}, and for fast algorithms for large-scale networks, see, e.g., \cite{Girvan2002,Clauset2004,Rosvall2008}. For datasets, see, e.g., SNAP network datasets \cite{snapnets} and SNAP biomedical datasets \cite{biosnapnets}. 

The soft graph clustering (SGC) problem, (also known as fuzzy graph clustering), on the other hand, allows clusters to have overlaps.  
A vertex may be a member of more than one cluster. There are numerous applications of SGC, such as: brain research, social network research, natural language processing, citation, and collaboration networks, and so on. A precise problem definition of the SGC varies and is dependent on the application, and sometimes it may not be possible to provide a precise problem definition. 

The subject of study in this paper considers the combinatorial optimisation problem where we are required to determine: 1) the composition of each of the clusters; 2) for each vertex that belongs to more than one cluster, how the membership is distributed amongst the clusters (we denote this by $x_{ic}$, for Vertex $i$ in Cluster $c$, hence $\sum_c x_{ic} = 1$ for all $i \in V$). We consider the case that an equal balance of the cluster total vertex memberships is desirable, and that not all vertices are required to be in a cluster. We consider two equally important objectives: 1) to minimize the sum of inter-cluster edge weights (cut across clusters); and 2) to maximize the sum of intra-cluster edge weights (cluster association). 
\subsection{Literature review} 

There are a number of existing soft clustering algorithms, each designed to suit different applications 
see, e.g., CFinder of \cite{Palla2005}, (see also \cite{Derenyi2005}, 
\cite{Adamcsek2006}), the MaxMax Algorithm of \cite{Hope2013}, the WATSET methods of \cite{Ustalov2018} for NLP,  the Chinese Whisper method of \cite{Biemann2006},  Betweenness-based method of \cite{Pinney2006}, and the Purifying and Filtering the Coupling Matrix approach of \cite{Liu2016}. 
Of these methods, \cite{Biemann2006},  \cite{Pinney2006}, \cite{Liu2016}, and \cite{Ustalov2018} are designed for unweighted graphs only (i.e., graphs with unit edge weight).The MaxMax Algorithm is designed for weighted undirected graphs. For unweighted graphs, however, it will return a trivial solution--each connected component of the graph will be a cluster. The CFinder is based on the finding of $k$-clique neighbourhoods. The mixed-integer linear programming method we propose in this paper is able to accomodate both weighted and unweighted graphs, with a small modification required for the latter. In the preliminary experiments section, we will compare and contrast the different methods.

We are not aware of any mixed-integer linear programming (MILP) models for the SGC problems. There are, however, MILP models for other graph clustering, machine learning, and data classification problems. 
The article \cite{Bertsimas2007} presents MILP formulations for classification and regression. The idea for classification, e.g., is to partition Class 1 points into $K$ disjoint subsets by finding the hyperplanes that describe the partitioning polyhedrons such that no Class 0 points can be expressed as a convex combination of the Class 1 points in each partition. 
In general clustering problems, \cite{Saglam2006} proposes a MILP formulation where one wishes to partition a set of data set into $k$ (a predetermined number of) clusters. The objective is to minimize the maximum diameter of the generated clusters in order to obtain evenly compact clusters. Essentially the method is an IP-based heuristic method, with some variables fixed by the solution of maximal independent set of size $k$, where each member of this set is a seed member in the $k$ clusters. The IP model (which is in fact a bilinear model, but linearized using standard linearization strategies) is then solved to obtain an optimal solution to the general clustering problem. 
Other MILP-based work can be found in, e.g., \cite{Gilpin2013} and \cite{Ye2007} for hierarchical clustering. The latter presents an application in recommendation systems.  
In Clique Covering Problem (CCP), an NP-hard combinatorial optimisation problem where an undirected graph is to be partitioned to form complete subgraphs, \cite{Miyauchi2018} proposes a compact ILP formulation for a relaxed problem, as well as a post-optimization repair procedure and a proof of optimality for the final solution to the original problem.

\subsection{Contribution of the paper} 

As far as we aware, this paper is the first to propose a methodology for SGC that i) deals with undirected graphs with general integer edge weights $w_e \geq 0$ and truely takes the values of $w_e$ into the optimization process in the way that the larger the value is, the more favourably it will be considered, and at the same time, also deals with unweighted graphs (i.e., graphs with unit edge weight); ii) simultaneously allocates membership proportion (the $x_{ic}$ values) for vertices that lie in multiple clusters; and iii) it enforces an equal balance of the clusters that the sum of vertex memberships over all clusters are roughly the same. We propose an approach that is based on a polynomial-size MILP model, beginning with a small value for $K$--we enforce that the graph has at least $K$ clusters. One can apply an adaptive approach to find the best value of $K$ iteratively, but the focus of this research is to solve an instantaneous SGC problem with a given $K$.

The method of \cite{Palla2005} requires obtaining $E' = \{e \in E \ | \ w_e > w^*\}$, and the graph is subsequently clustered by finding $\kappa$-clique neighbourhoods on an unweighted graph $H = (V,E')$. Our method does not require the finding of $\kappa$-cliques, and it takes the values of $w_e$ into account during optimization. Comparing with the method of \cite{Hope2013} (MaxMax), for unweighted graphs, if the graph is connected, then MaxMax will produce only one cluster which is the entire graph. 
Our method, however, can deal with unweighted graphs by converting them into weighted ones via a simple transformation.  

In Section \ref{Sec:BasicMILP}, we present our basic MILP model by considering a number of standard requirements for the SGCP. In Section \ref{Sec:connectivity}, we discuss our strategies for graph connectivity. In Section \ref{Sec:Obj}, we discuss two objectives: 1) minimizing the total inter-cluster cut, and 2) maximize the total intra-cluster association. In Section \ref{Sec:NumRes}, we present preliminary numerical results. 
We then conclude our findings and discuss future research directions in Section \ref{Sec:Conc}.

\section{A mixed-integer linear programming formulation} 
\label{Sec:MILP}


\subsection{The basic model}
\label{Sec:BasicMILP}

We first introduce the notation used in this paper. Let: 
\bi
\item 
$A = \{a_{ij} \in \{0,1\} \ | \ i,j \in V\}$ be the adjacency matrix of $G$; 
\item 
$M_w = \max\{w_e \ | \ e \in E\}$ the maximum edge weight; 
\item 
$K$ be the number of clusters; 
\item 
$\mC = \{c_1\ldots,c_K\}$ be the set of clusters; 
\item 
$y_{i,c} \in \{0,1\}$ be a binary decision variable with $y_{i,c}=1$ indicating Vertex $i$ is a member of Cluster $c$; 
\item 
$x_{i,c} \in [0,1]$ be a continuous decision variable indicating the membership of Vertex $i$ in Cluster $c$; 
\item 
$\kappa (c_1,c_2)$ be the cut between clusters $c_1$ and $c_2$; 
\item 
$0 < \mu < 1$ a predetermined minimum membership if a vertex is a member of a cluster; 
\item 
$0 < \delta <1$ a predetermined tolerance equal balance of cluster membership;  
\item 
$0 < \nu < 1$ a predetermined maximum overlap factor. 
\ei 
Now we introduce the constraints. 
First, we have a set of {\em Membership Constraints}.  
The membership of Vertex $i$ in Cluster $c$ can only be non-zero if it is a member of $c$, and when it is, the membership must be no less than a predetermined value. 
\begin{alignat}{5}
x_{i,c}  & \leq & y_{i,c}, 
	& \quad \forall i \in V, \ \forall c \in \mC 
		\label{LogicConstraint1} 	\\
x_{i,c}  & \geq & \ \mu  y_{i,c}, 
	& \quad \forall i \in V, \ \forall c \in \mC 
		\label{MinMembership}	
\end{alignat} 

Let $L_i$, for each $i \in V$, be an auxiliary binary variable with $L_i=1$ indicating $i$ is in at least one of the clusters and $L_i=0$ otherwise. We require that the sum of memberships for any vertex over all clusters is exactly 1 if the vertex is a member of at least one cluster and 0 otherwise. 
\begin{alignat}{5}
y_{i,c}  & \leq & L_i, 
	& \quad \forall i \in V, \ \forall c \in \mC 
		\label{LogicConstraint2}	\\	
L_i & \leq & \ \sum_{c \in \mC} y_{i,c},  
	& \quad \forall i \in V
		\label{LogicConstraint3}	\\
\sum_{c \in \mC} x_{i,c} & = & \ L_i, 
	& \quad \forall i \in V 
		\label{SumOfMembershipEachVertex}
\end{alignat} 
We consider an {\em Equal Balance Requirement} where the sum of memberships in all clusters are ``roughly'' the same. As far as we aware, this is the first method that considers such a requirement. This requirement can be modelled as below. 
\begin{alignat}{5}
(1-\delta) \sum_{i \in V} x_{i,c_1} \leq \sum_{i \in V} x_{i,c_2} \leq (1+ \delta) \sum_{i \in V} x_{i,c_1}, \notag \\
	\quad \forall c_1,c_2 \in \mC, c_1 \neq c_2
 		\label{BalancedMembershipConstraint}
\end{alignat} 
Next, we consider the {\em Overlap Cardinality Constraints}.   
Let $t^i_{c_1,c _2}$ be auxiliary binary decision variable such that $t^i_{c_1,c _2} = 1$ if and only if $i$ is a member of both clusters $c_1$, $c_2$, i.e., $y_{i,c_1} = y_{i,c_2} = 1$. We have that:  
\begin{alignat}{5}
y_{i,c_1} + y_{i,c_2} & \leq  &\  t^i_{c_1,c _2} + 1, 
	&\ \forall i \in V,  
	c_1,c_2 \in \mC, c_1 \neq c_2 
		\label{OverlapsCardinalityCosntraints1} \\ 
 t^i_{c_1,c _2}  & \leq &   y_{i,c_1} , 
 	& \   \forall i \in V, 
	c_1,c_2 \in \mC,  c_1 \neq c_2 
		\label{OverlapsCardinalityCosntraints2} \\
 t^i_{c_1,c _2}  & \leq &   y_{i,c_2} , 
 	& \  \forall i \in V, 
	c_1,c_2 \in \mC, c_1 \neq c_2 
		\label{OverlapsCardinalityCosntraints3} 
\end{alignat} 
Given any pairs of clusters $c_1,c_2 \in \mC$, with $c_1 \neq c_2$, the number of vertices that are in both clusters cannot be larger than a predetermined fraction, $\nu$, of the cardinality of either of the clusters. 
\begin{alignat}{5}
\sum_{i \in V}  t^i_{c_1,c _2}  & \leq &\  \nu \sum_{i \in V}  y_{i,c_1} 
	& \quad  \forall c_1,c_2 \in \mC, \ c_1 \neq c_2 
		\label{OverlapsCardinalityCosntraints4} \\
\sum_{i \in V}  t^i_{c_1,c _2}  & \leq & \ \nu \sum_{i \in V}  y_{i,c_2} 
	& \quad  \forall c_1,c_2 \in \mC, \ c_1 \neq c_2 
			\label{OverlapsCardinalityCosntraints5}
\end{alignat} 
To calculate the {\em Inter-cluster Cuts} between $c_1$ and $c_2$, we require auxiliary binary variables and nonlinear terms. 
First, let $\eta^{i,j}_{c_1,c_2}$ be a binary variable such that $\eta^{i,j}_{c_1,c_2} = 1$ if both of $i,j$ are in the intersection of clusters $c_1$ and $c_2$.    
\begin{alignat}{5}
t^i_{c_1,c_1} + t^j_{c_1,c_1} & \leq & \ \eta^{i,j}_{c_1,c_2} + 1, \ &    \forall i \neq j \in V, \ c_1 \neq c_2 \in \mC  \\
\eta^{i,j}_{c_1,c_2}  & \leq & t^i_{c_1,c_1}, \  &    \forall i \neq j \in V, \ c_1 \neq c_2 \in \mC  \\
\eta^{i,j}_{c_1,c_2}  & \leq & t^j_{c_1,c_1}, \  &    \forall i \neq j \in V, \ c_1 \neq c_2 \in \mC
\end{alignat} 
We then use a binary variable $s^e_{c_1,c_2}$ to indicate the existence of an edge (cut) $e = (i,j)$ (i.e., $a_{ij} = 1$) with $i$ in $c_1$ and $j$ in $c_2$, but not both in the intersection of $c_1$ and $c_2$ (otherwise the ``cut'' should not be counted). The constraints are as below. 
For each pair of distinct vertices $i,j \in V, \ i \neq j$ and each pair of distinct clusters $c_1,c_2 \in \mC, \ c_1 \neq c_2$, we have that: 
\begin{alignat}{5}
y_{i,c_1} + y_{j,c_2} + a_{ij} + (1-\eta^{i,j}_{c_1,c_2}) & \leq & s^{e}_{c_1,c_2} + 3
	 	\label{CutCalculation1} \\
s^{e}_{c_1,c_2} & \leq  & y_{i,c_1}
		\label{CutCalculation2} \\
s^{e}_{c_1,c_2} & \leq & y_{j,c_2}
			\label{CutCalculation3} \\
s^{e}_{c_1,c_2} & \leq & a_{ij}
			\label{CutCalculation4} \\
s^{e}_{c_1,c_2} & \leq & (1- \eta^{i,j}_{c_1,c_2}) \label{CutCalculation5} 
\end{alignat} 
%
%
Now, the cut between two distinct vertices $i,j \in V$ across two distinct clusters $c_1,c_2 \in \mC$, if edge $e=(i,j) \in E$ exists, is defined by: 
%
$w_{e}  v^{e}_{c_1,c_2}$, 
%
for   
\begin{alignat}{5}
v^{e}_{c_1,c_2}= \left(x_{i,c_1} + x_{j,c_2} \right) s^{e}_{c_1,c_2} \label{CutCalculation6} 
\end{alignat}
The terms $x_{i,c_1} s^{e}_{c_1,c_2}$ and $x_{j,c_2} s^{e}_{c_1,c_2}$ are bilinear, and can be linearized by introducing auxiliary non-negative continuous variables $\tau^{e,i}_{c_1,c_2} \geq 0$ and $\tau^{e,j}_{c_1,c_2} \geq 0$ and the following constraints. 
For each $e = (i,j) \in E,  \ c_1, c_2 \in \mC, \ c_1 \neq c_2 $, we have that: 
\begin{alignat}{5}
\tau^{e,i}_{c_1,c_2} & \leq & \  x_{i,c_1}
		\label{Linearization1} \\
\tau^{e,i}_{c_1,c_2} & \leq & \  s^{e}_{c_1,c_2} 
		\label{Linearization2} \\
\tau^{e,i}_{c_1,c_2} & \geq & \ -1 + s^{e}_{c_1,c_2} + x_{i,c_1}
		\label{Linearization3} \\
\tau^{e,j}_{c_1,c_2} & \leq & \  x_{j,c_2}
		\label{Linearization4} \\
\tau^{e,j}_{c_1,c_2} & \leq & \ s^{e}_{c_1,c_2}
		\label{Linearization5} \\
\tau^{e,j}_{c_1,c_2} & \geq & \ -1 + s^{e}_{c_1,c_2} + x_{j,c_2}
		\label{Linearization6} 
\end{alignat} 
The cut $\kappa(c_1,c_2)$ is calculated by the following linear term. 
\begin{alignat}{5}
\kappa(c_1,c_2) = \sum_{(i,j) \in E} w_{i,j}(\tau^{e,i}_{c_1,c_2} + \tau^{e,j}_{c_1,c_2} ) \label{kappaValue}
\end{alignat} 

Now, we consider the {\em Intra-cluster Association} calculations.  
Let $z^c_{i,j}$ be an auxiliary binary variable with $z^c_{i,j} = 1$ if $i,j$ are both in $c$ and that the edge $e = (i,j)$ exists in $E$. 
The constraints are give by:   
\begin{alignat}{5}
y_{i,c} + y_{j,c} + a_{ij} & \leq & z^c_{i,j} +2,  
	& \quad (i,j) \in E, \ \forall c \in \mC  
		\label{AssociationConstraint1}\\
z^c_{i,j} & \leq & y_{i,c}, 
	& \quad (i,j) \in E, \ \forall c \in \mC 
		\label{AssociationConstraint2} \\
z^c_{i,j} & \leq & y_{j,c}, 
	& \quad (i,j) \in E,\ \forall c \in \mC  
		\label{AssociationConstraint3}\\
z^c_{i,j} & \leq & a_{ij}, 
	& \quad (i,j) \in E, \ \forall c \in \mC 
		\label{AssociationConstraint4}
\end{alignat} 
The intra-cluster association of $c \in \mC$ is given by 
\begin{alignat}{5}
A(c) = \sum_{(i,j) \in E} w_{i,j} (x_{i,c} + x_{j,c})z^c_{i,j} \notag
\end{alignat} 
We linearize the association using auxiliary continuous variables $\pi^{c,i}_{i,j}$ and $\pi^{c,j}_{i,j}$ to capture the memberships of two vertices in the same cluster should an edge exists between them. I.e., when $z^c_{i,j} = 1$, $\pi^{c,i}_{i,j} =x_{i,c}$, otherwise, $\pi^{c,i}_{i,j}=0$. (Similarly for $\pi^{c,j}_{i,j}$). 
The constraints are as below. 
\begin{alignat}{5}
\pi^{c,i}_{i,j} & \leq & \  x_{i,c}, 
	& \quad \forall e = (i,j) \in E, c \in \mC 
		\label{Linearization11} \\
\pi^{c,i}_{i,j} & \leq & \  z^c_{i,j}, 
	& \quad \forall e = (i,j) \in E, c \in \mC 
		\label{Linearization12} \\
\pi^{c,i}_{i,j} & \geq & \ -1 + z^c_{i,j}  + x_{i,c}, 
	& \quad \forall e = (i,j) \in E, c \in \mC 
		\label{Linearization13} \\
\pi^{c,j}_{i,j} & \leq & \  x_{j,c}, 
	& \quad \forall e = (i,j) \in E, c \in \mC 
		\label{Linearization14} \\
\pi^{c,j}_{i,j} & \leq & \  z^c_{i,j} 
	& \quad \forall e = (i,j) \in E, c \in \mC 
		\label{Linearization15} \\
\pi^{c,j}_{i,j} & \geq & \ -1 + z^c_{i,j}  + x_{j,c}, 
	& \quad \forall e = (i,j) \in E, c \in \mC 
		\label{Linearization16} 
\end{alignat} 
The association within a cluster $c$, denoted by $A(c)$, is calculated as follows: 
\begin{alignat}{5}
A(c) = \sum_{(i,j) \in E} w_{i,j} (\pi^{c,i}_{i,j} + \pi^{c,j}_{i,j} ) \label{AssoValue}
\end{alignat}

\subsection{Cluster connectivity} 
\label{Sec:connectivity}

Clusters are required to be connected, it is likely that this can only be achieved with exponentially many variables and solved using a branch-and-price approach, or with exponentially many constraints and solved using a branch-and-cut approach.  
Our approach here is not an exact one, as we wish to keep the formulation compact (i.e., polynomial in size). We derive a number of constraints that capture a few conditions for graph connectivity that are necessary but not sufficient. When there is a connectivity violation,  violation elimination constraints can be added in a lazy fashion. ({\em Lazy constraints} is a technical term in integer programming--hard (and often exponentially many) constraints are relaxed, and are only added when the current integer optimal solution to the relaxed problem violates them--usually only a very small number of them are violated--the problem is then re-optimized, and the procedure recurs until there is no more violated hard constraints). In our preliminary test where we used randomly generated undirected graphs (see Preliminary Numerical Results section), most of the problem instances produced connected clusters. 

First of all, the number of edges in a connected undirected graph cannot be smaller than the cardinality of the graph minus one. We define $\gamma^c_{i,j} \in \{0,1\}$ to be a {\em span variable} that can only be one when $i$ and $j$ are both in Cluster $c$ and that the edge $(i,j)$ exists in $E$. We have that: 
\begin{alignat}{5}
\gamma^c_{i,j} & \leq & \ a_{ij}, 
	& \quad (i,j) \in E, \ \forall c \in \mC 
		\label{ConnectivityConstraint1} \\
\gamma^c_{i,j} & \leq & \ y_{i,c}, 
	& \quad (i,j) \in E, \ \forall c \in \mC 
		\label{ConnectivityConstraint2} \\
\gamma^c_{i,j} & \leq &\ y_{j,c}, 
	& \quad (i,j) \in E,\ \forall c \in \mC  
		\label{ConnectivityConstraint3}
\end{alignat} 
and that 
\begin{alignat}{5}
\sum_{i \in V} y_{i,c} -1 & \leq & \sum_{(i,j) \in E} \gamma^c_{i,j}, & \quad  \forall c \in \mC \label{ConnectivityConstraint4} 
\end{alignat} 

We also require that all vertices in each cluster must be connected to at least one other vertex in the same cluster. 
\begin{alignat}{5}
y_{i,c} \leq \sum_{j \in V \setminus \{i\}\ : \ (i,j) \in E} z^c_{ij}, & \quad \forall i \in V, \ \forall c \in \mC \label{ConnectivityConstraint5}
\end{alignat} 
However, Constraints (\ref{ConnectivityConstraint1})--(\ref{ConnectivityConstraint5}) are not enough to eliminate multiple loops within a cluster. A cluster may contain vertices $\{1,2,3,4,5,6\}$, but the constraints cannot eliminate the formation of subgraphs  $\{1,2,3\}$ and $\{4,5,6\}$ within the cluster. 

Therefore, we borrowed the time constraints idea for Asymmetric Travelling Salesman Problem (ATSP) subtour elimination (\cite{Miller1960}). Let $t_i \geq 0$, $i \in V$, be a decision variable indicating the ``time of arrival'' at Vertex $i$. 
Suppose $\{1,2,3\}$ is a strict subset of a cluster $c$, to prevent the relevant span variables from forming a loop, we require that $\gamma^c_{i,j}=1$ if and only if $t_j = t_i +1$, for each pair of distinct $i, j \in V$, $i \neq j$ and each $c \in \mC$. We have that: 
\begin{alignat}{5}
-(|V|+1)(1-\gamma^c_{i,j})+1 & \leq t_j - t_i 
\label{timeConstr1} \\
t_j - t_i  & \leq 1 + |V| (1-\gamma^c_{i,j})
\label{timeConstr2}
\end{alignat} 
E.g., consider the vertices $\{1,2,3\}$ a strict subset of $c$, a loop with edges $(1,2), (2,3), (1,3)$ a violation of the time constraints, as $t_3$ cannot be equal to $t_1 + 1$ and $t_1+2$ simultaneously. 
Constraints (\ref{timeConstr1}) and (\ref{timeConstr2}) can eliminate certain types of loops formed by the span variables, but not all. In any case,  adding them will give the span variables a better chance at making a full span in a cluster and thus the connectivity. 

Notice that the time constraints do not cut off feasible solutions due to the fact that even if a variable $\gamma^c_{i,j}$ is forced to be 0, there is no impact on the values of $y_{ic}$ and $y_{jc}$, as (\ref{ConnectivityConstraint1})--(\ref{ConnectivityConstraint3}) do not induce a bi-conditional relation. In our preliminary numerical experiments, the time constraints are expensive to implement, and often we do not have disconnected clusters with just (\ref{ConnectivityConstraint1})--(\ref{ConnectivityConstraint5}) alone. 

When cluster size $K$ is a hard constraint, and there exist a disconnected cluster, in the case of maximizing total association, e.g., one can consider adding the following cut in a lazy fashion and re-optimize. Let $\mI^*$ be the set of $y$-variables with a value of 1 in the optimal solution, we add: 
\begin{alignat}{5}
\sum_{(i,c) \in \mI} y_{i,c} \leq |\mI^*|-1. 
\end{alignat}  

\subsection{The objective function} 
\label{Sec:Obj}

Two commonly considered objective functions are: 1) minimize the sum of inter-cluster cuts and 2) maximize the sum of intra-cluster association given by (\ref{ObjFn1}) and (\ref{ObjFn2}) below respectively. 
\begin{alignat}{5}
\min z = \sum_{\forall c_1,c_2 \in \mC, \ c_1 \neq c_2} 
	\kappa(c_1,c_2) \label{ObjFn1}
\end{alignat} 
\begin{alignat}{5}
\max w = \sum_{\forall c \in \mC} 
	A(c) \label{ObjFn2}
\end{alignat} 
Notice that it is necessary to enforce a minimum cluster size, otherwise a trivial optimal solution to (\ref{ObjFn1}) is to assign no vertices to any clusters. Thus we have: 
\begin{alignat}{5}
\sum_{c \in \mC} \sum_{i \in V} y_{i,c} \geq \sigma |V|,  \label{MinClusterSize}
\end{alignat} 
for $0 < \sigma < 1$ a predetermined value. 


The two objectives should be considered simultaneously. Some may consider minimizing the sum of ratio of inter-cluster cuts and intra-cluster associations over all pairs of distinct clusters. 
\begin{alignat}{5}
\min z = \sum_{\forall c_1,c_2 \in \mC, \ c_1 \neq c_2} 
	\dfrac{\kappa(c_1,c_2)}{A(c_1)+A(c_2)} \label{ObjFn3}
\end{alignat} 
Diagram (\ref{Dia:graph1}) below shows how the sum of ratios of inter-cluster cuts and intra-cluster associations changes. The first data point on the left is obtained when we optimize (\ref{ObjFn1}) and set $w^1$ to be the value of $\sum_{\forall c \in \mC} A(c)$ in the optimal solution. We can see that the total intra-cluster association is also small, and so is the value of (\ref{ObjFn3}). The last data point on the right is obtained when we optimize (\ref{ObjFn2}) and we set $w^2$ to be the optimal objective value. We then minimize (\ref{ObjFn1}) with $\sum_{\forall c \in \mC} A(c) \geq \ell_j$, for $\ell_j = \frac{j}{10}  (w^2 - w^1)$, $j = 1,\ldots,10$.

\begin{figure}[ht]
\vskip 0.2in
\begin{center}
\centerline{\includegraphics[width=6cm]{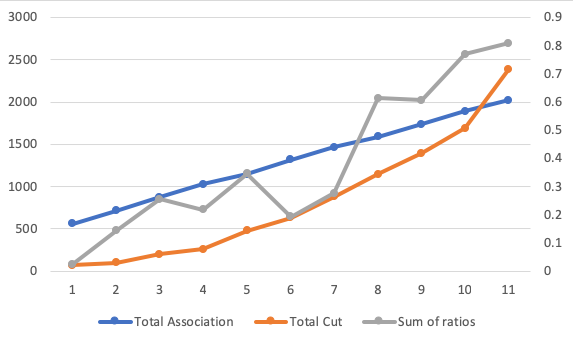}}
\caption{Changes of Total Association, Total Cut, and Sum of Ratios.}
\label{Dia:graph1}
\end{center}
\vskip -0.2in
\end{figure}

One can, however, consider recent advances in Bi-objective Integer Programming (see, e.g., \cite{Dai2018}). In fact, different applications of the SGC may have a different objective function that describes the SGC better. Besides, the structure of the graph must be taken into consideration in determining what the most appropriate objective function is. One may therefore consider applying machine learning to automate the finding of an appropriate objective function.

\subsection{Edge weight transformation for unweighted graphs} 
\label{edgeWeightConversion}

For undirected graphs with $w_e = 1$, for all $e = (i,j) \in E$, the MILP does not work very well because it cannot distinguish between an edge in a sparse neighbourhood with one in a dense neighbourhood. To give favour to edges in a dense neighbourhood, we obtain new edge weights by calculating $w'_e = 1 + | \{ k  \in V \ : \ (i,k),(j,k) \in E, \ k \neq i,j \} |$, i.e., the new edge weight of $e$ will be one plus the number of vertices that are connected to both of the end nodes of $e$. 

\section{Preliminary numerical results}
\label{Sec:NumRes}

We compared our method with {\em CFinder} \cite{Palla2005,Derenyi2005,Adamcsek2006} and MaxMax Algorithm \cite{Hope2013}. We used a 21-vertex instance of weighted graph. We can see that the clusters of CFinder are the $3$-clique neighbourhoods, and therefore some of the edges with heavy weight are not included, e.g., $(6,7)$, $(9,10)$, $(17,20)$. 

\begin{figure}[ht]
\vskip 0.2in
\begin{center}
\centerline{\includegraphics[width=5.8cm]{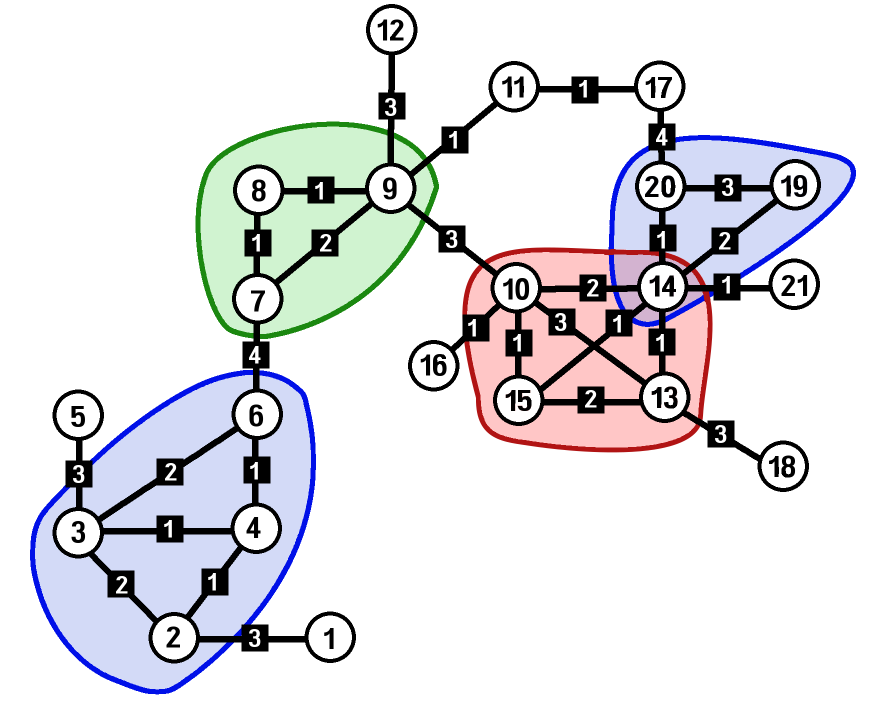}}
\caption{CFinder solution.}
\label{Dia:graph2}
\end{center}
\vskip -0.2in
\end{figure}

The MaxMax solution, however, have these heavy-weight edges covered. However, for unweighted graphs, MaxMax will cluster the graph by connected components, so if the graph is connected, then there will be only one cluster. 

\begin{figure}[ht]
\vskip 0.2in
\begin{center}
\centerline{\includegraphics[width=5.8cm]{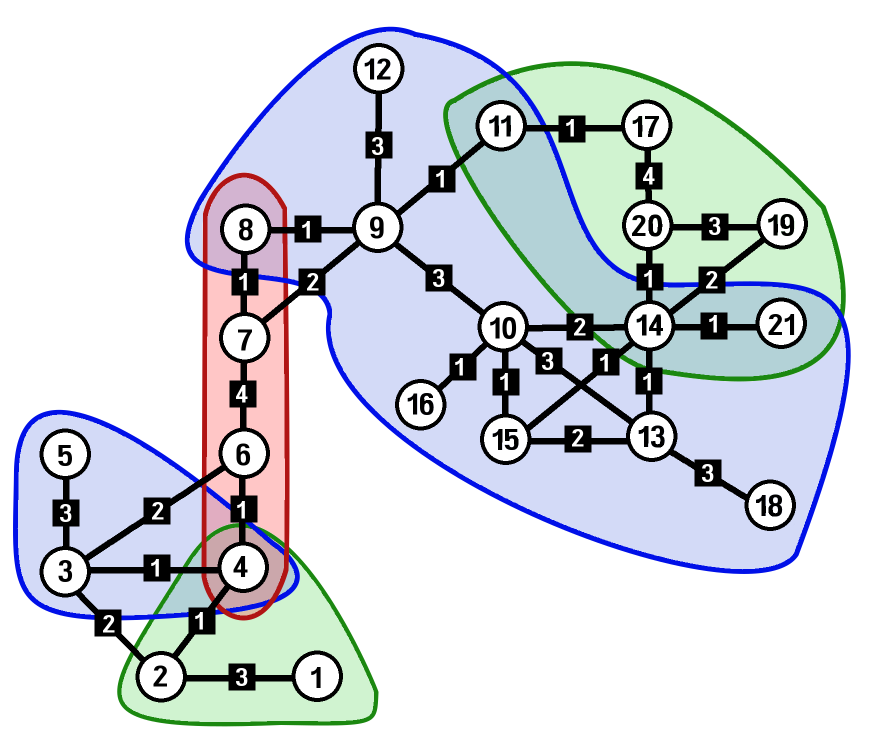}}
\caption{MaxMax solution.}
\label{Dia:graph3}
\end{center}
\vskip -0.2in
\end{figure}

The MILP solution produces not only the clusters, but also the membership proportion for each vertex that belongs to more than one cluster. None of MaxMax or CFinder provides this information. One can see clearly that the Min Cut solution can also produce a relatively balanced set of clusters in terms of total weighted memberships of the clusters when the clusters are connected. None of the existing SGC methods considered such a constraint. 
\begin{figure}[ht]
\vskip 0.2in
\begin{center}
\centerline{\includegraphics[width=5.8cm]{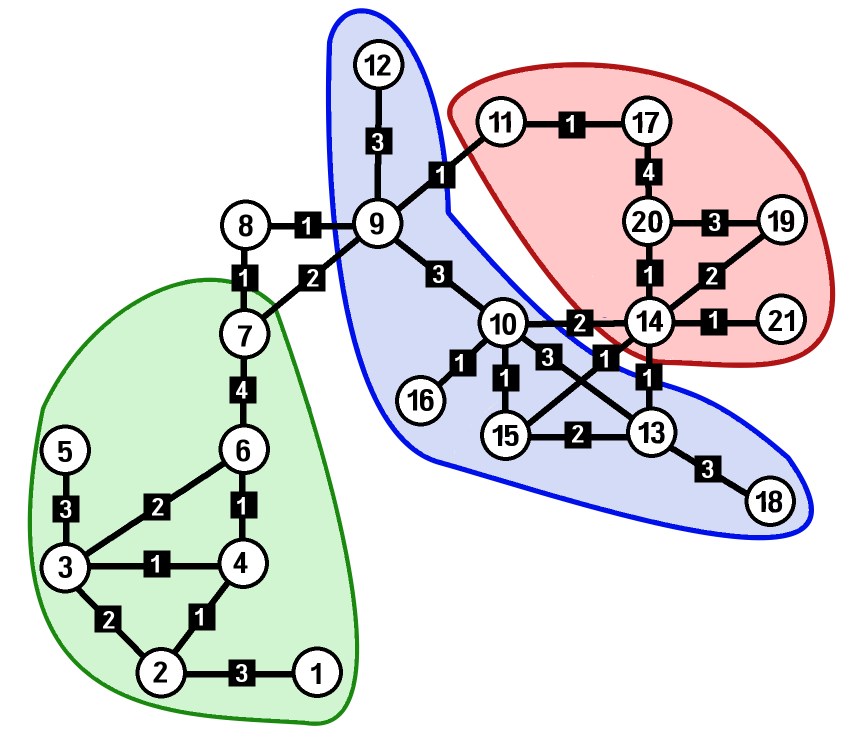}}
\caption{Minimium Total Cut solution.}
\label{Dia:graph4}
\end{center}
\vskip -0.2in
\end{figure}

In the Max Association solution below, unfortunately, because the MILP enforced three clusters, and the blue cluster is not connected, we have four clusters instead, so the equal balance cannot be guarantee in this case. In any case, the four clusters demonstrated the four strongest connections by $w_e$. As for the membership proportions, take Vertex 9 as an example, since there are more heavily weighted links in the pink cluster rather than the green cluster, 0.9 of its membership is in the former, and 0.1 in the latter.  

\begin{figure}[ht]
\vskip 0.2in
\begin{center}
\centerline{\includegraphics[width=5.8cm]{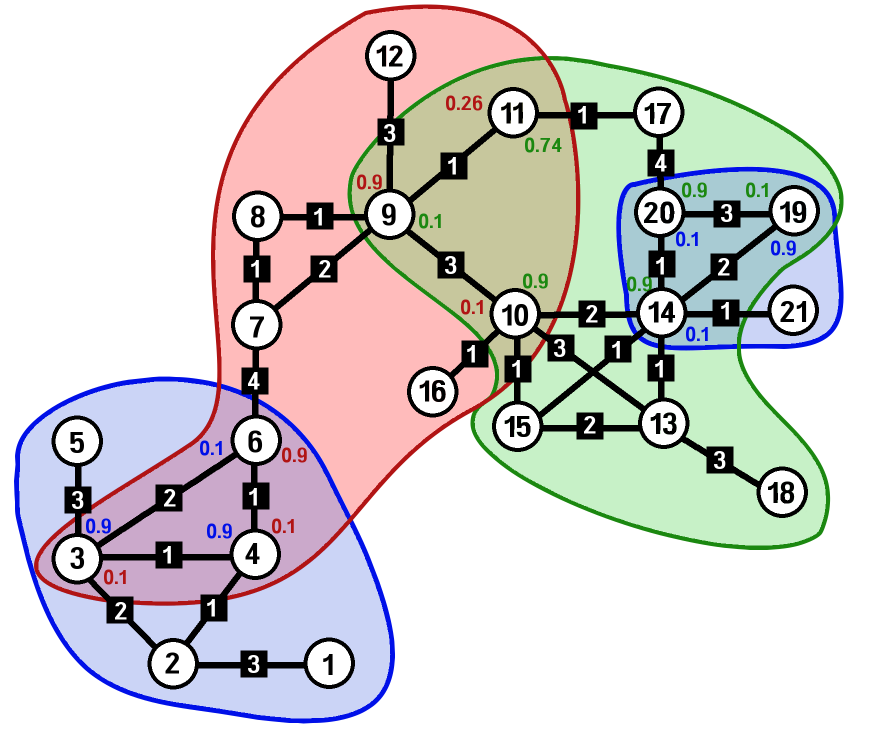}}
\caption{Maximum Total Association solution.}
\label{Dia:graph5}
\end{center}
\vskip -0.2in
\end{figure}

\subsection{KKI instances} 

In this section, we applied our method on some KKI instances from 
\url{https://github.com/shiruipan/graph_datasets}. Since the KKI instances are unweighted, we perform the edge weight transformation described in Section \ref{edgeWeightConversion}. 

In Table \ref{table1a}, column {\em Opt} presents the time taken (in seconds) for proving optimality, and the column {\em Gap} presents the gap between the MIP relative gap, (i.e., the gap between the best integer objective and the objective of the best active node on the branch-and-bound tree), for instances that are not solved to optimality within the given time limit (10 minutes). The column {\em r} is the ratio of total inter-cluster cut over total intra-cluster association.  
We used IBM ILOG CPLEX Optimization Studio v12.8 \cite{Cplex} for solving the MILPs.

\begin{table}[t]
\caption{Computational results for KKI Instances with 20-50 vertices.  
}
\label{table1a}
\vskip 0.15in
\begin{center}
\begin{small}
\begin{sc}
\resizebox{\columnwidth}{!}{
\begin{tabular}{lcccccc}
\toprule
 &    \multicolumn{3}{c}{Min Cut} &    \multicolumn{3}{c}{Max Association}  \\
Data set 	& Opt & Gap & $r$ 	& Opt & Gap &  $r$   \\
\midrule
1019436-35& 3.66& 0.00& 0.00& 48.57& 0.00& 0.02 \\
1541812-21& 1.16& 0.00& 0.00& 8.64& 0.00& 0.23 \\
1577042-49& 39.28& 0.00& 0.00& 118.26& 0.00& 0.03 \\
1638334-34& 5.88& 0.00& 0.00& 94.91& 0.00& 0.04 \\
1735881-35& 1.68& 0.00& 0.00& 37.56& 0.00& 0.13 \\
1779922-28& 2.24& 0.00& 0.00& 107.93& 0.00& 0.16 \\
2371032-46& 34.49& 0.00& 0.00& 600.07& 0.03& 0.16 \\
2558999-28& 0.82& 0.00& 0.00& 19.56& 0.00& 0.17 \\
2601925-29& 3.62& 0.00& 0.00& 62.78& 0.00& 0.14 \\
2618929-20& 3.09& 0.00& 0.00& 31.56& 0.00& 0.46 \\
2768273-23& 10.57& 0.00& 0.09& 30.90& 0.00& 0.09 \\
2903997-44& 32.75& 0.00& 0.00& 500.42& 0.00& 0.22 \\
2917777-44& 22.55& 0.00& 0.00& 110.44& 0.00& 0.01 \\
3310328-47& 19.49& 0.00& 0.00& 208.22& 0.00& 0.07 \\
3611827-39& 75.61& 0.00& 0.00& 600.04& 0.03& 0.20 \\
3713230-20& 11.23& 0.00& 0.00& 23.83& 0.00& 0.15 \\
3902469-45& 34.21& 0.00& 0.00& 114.55& 0.00& 0.13 \\
3972472-31& 12.07& 0.00& 0.00& 101.97& 0.00& 0.10 \\
4104523-25& 6.11& 0.00& 0.00& 30.87& 0.00& 0.18 \\
4275075-28& 274.13& 0.00& 0.00& 600.05& 0.00& 0.15 \\
4362730-38& 28.72& 0.00& 0.00& 64.47& 0.00& 0.04 \\
5216908-30& 37.06& 0.00& 0.00& 113.23& 0.00& 0.14 \\
6346605-27& 600.05& 0.93& 0.00& 600.05& 0.00& 0.32 \\
7129258-34& 4.72& 0.00& 0.00& 70.97& 0.00& 0.17 \\
7415617-28& 5.49& 0.00& 0.00& 106.67& 0.00& 0.30 \\
7774305-20& 0.95& 0.00& 0.00& 13.82& 0.00& 0.10 \\
8263351-42& 63.80& 0.00& 0.00& 208.89& 0.00& 0.11 \\ \hline 
\end{tabular}
}
\end{sc}
\end{small}
\end{center}
\vskip -0.1in
\end{table}

\subsection{Instances from a random problem generator}

We have generated 10 problem classes with 5 instances for each problem class. The problem instances were generated using a random problem generator where number of vertices on the undirected graph, ($|V|$), density of graph $\frac{|E|}{2|V|(|V|-1)}$, and an upper bound on the edge weight $M_w$ are taken as user input. The edge weights are generated by each given a number chosen uniformly randomly between $[1,M_w]$. In Table \ref{table1c}, the names of the problem classes are in the form of: N followed by the value of $|V|$, d followed by the density of $G$, and then $M$ followed by the value of $M_w$. 
The column {\em CON} is the percentage of clusters that are connected in each problem class.  
In the columns where two numbers were given for each problem class, the first row presents the average value over 5 problem instances, and the second row presents the value of the standard deviation. The bracket underneath each problem class  indicates the number of instances that are solved to optimality versus the number of instances that are not.

\subsection{Preliminary testing}

We experimented with instances with maximum edge weight $M_w = 50$ and $100$, $K=3$ clusters, and  $\sigma = 0.7$.  From the results of Table \ref{table1c}, we can see that the computation time grows exponentially as the size of the problem grows, or the density of the graph grows. For problem instances that are not solved to optimality, the MIP gaps are large. 
However in all problem instances, feasible solutions are found reasonably quickly.  We can see that the computation time for maximizing total intra-cluster association is substantially longer than minimizing total inter-cluster cuts. For problem instances that are not solved to optimality, however, the former has a smaller MIP gap whilst the latter can solve larger problem instances. With an objective of maximizing total association, we did not include Constraint (\ref{MinClusterSize}) as the objective will drive as much edges used as possible. We cannot obtain a conclusive remark on the effects of $M_w$ in terms of computation time.

\begin{table}[t]
\caption{Computational results for problem instances with maximum edge weight $M_w = 50$ and 100, $K=3$ clusters, and at least $\sigma = 0.7$ vertices must be in at least one cluster.  
}
\label{table1c}
\vskip 0.15in
\begin{center}
\begin{small}
\begin{sc}
\resizebox{\columnwidth}{!}{
\begin{tabular}{lcccc|ccccc}
\toprule 
 &    \multicolumn{4}{c}{Min Cut} &     \multicolumn{5}{c}{Max Association}  \\
Data set 	& Opt & Gap  & r & Con & & Opt  & Gap & r & Con \\
\midrule
N15d015M50 & 0.939 & 0 & 0.215 & 100 & & 2.347 & 0 & 0.769 & 100 \\
(5/0) & 0.269 & 0 & 0.195 & &  (5/0) & 0.987 & 0 & 0.155 & \\
N15d025M50 & 2.012 & 0 & 0.219 & 100 & & 14.60 & 0 & 0.784 & 93.33 \\
(5/0) & 0.975 & 0 & 0.150 & &(5/0)  & 12.85 & 0 & 0.107 & \\
N15d05M50 & 78.46 & 0 & 0.393 & 100 & & 412.3 & 0.074 & 0.912 & 100 \\
(5/0) & 43.20 & 0 & 0.317 & & (4/1) & 85.48 & 0 & 0.198 & \\ \hline 
N20d015M50 & 2.612 & 0 & 0.053 & 100 & & 54.90 & 0 & 0.789 & 100 \\
(5/0) & 1.348 & 0 & 0.047 & & (5/0) & 32.11 & 0 & 0.193 & \\
N20d025M50 & 66.07 & 0 & 0.146 & 100 & & 252.2 & 0.067 & 0.876 & 100 \\
(5/0) & 51.50 & - & 0.110 & & (1/4) & 0 & 0.017 & 0.177 & \\
N20d05M50 & - & 0.686 & 0.484 & 100 && - & 0.229 & 1.069 & 100 \\
(0/5) & - & 0.176 & 0.152 & & (0/5) & - & 0.033 & 0.208 & \\ \hline 
N30d015M50 & 31.38 & 0.826 & 0.025 & 100 & & - & 0.044 & 0.817 & 100 \\
(4/1) & 11.07 & 0 & 0.048 & & (0/5) & - & 0.021 & 0.073 & \\
N30d025M50 & - & 0.969 & 0.224 & 100 & & - & 0.146 & 0.862 & 100 \\
(0/5) & - & 0.021 & 0.110 &  & (0/5) & - & 0.028 & 0.102 & \\ \hline 
N50d015M50 & - & 1 & 0.142 & 93.33  & & - & 0.176 & 0.937 & 100 \\
(0/5) & - & - & 0.095 & & (0/5) & - & 0.032 & 0.066 & \\
N50d025M50 & - & 1 & 0.683 & 93.33 & & - & 0.462 & 1.078 & 100 \\
(0/5) & - & 0 & 0.128 & & (0/5) & - & 0.049 & 0.078 & \\ \hline 
N15d015M100 & 0.540 & 0 & 0.089 & 80.00 & & 3.193 & 0 & 0.484 & 80.00 \\
(5/0) & 0.492 & 0 & 0.123 & & (5/0) & 5.069 & 0 & 0.476 & \\ 
N15d025M100 & 3.226 & 0 & 0.218 & 100 & & 44.01 & 0 & 0.886 & 100 \\ \hline
(5/0) & 2.417 & 0 & 0.143 & & (5/0) & 26.46 & 0 & 0.149 & \\
N15d05M100 & 35.86 & 0 & 0.359 & 100 & & 342.5 & 0.116 & 1.070 & 100 \\
(5/0) & 20.11 & 0 & 0.202 & & (3/2) & 95.53 & 0.021 & 0.184 & \\ \hline
N20d015M100 & 4.897 & 0 & 0.096 & 86.66 & & 41.86 & 0 & 0.789 & 73.33 \\
(5/0) & 3.398 & 0 & 0.105 & & (5/0) & 15.81 & 0 & 0.092 & \\
N20d025M100 & 71.12 & 0 & 0.102 & 100 & & 261.9 & 0.049 & 0.922 & 100 \\ 
(5/0) & 62.69 & 0 & 0.076 & & (1/4) & 0 & 0.018 & 0.146 & \\ 
N20d05M100 & - & 0.709 & 0.494 & 100 & & - & 0.210 & 1.052 & 100 \\
(0/5) & - & 0.064 & 0.155 & & (0/5) & - & 0.014 & 0.078 & \\ \hline
N30d015M100 & 46.06 & 0 & 0.006 & 93.33 & & 515.5 & 0.037 & 0.731 & 93.33 \\
(5/0) & 70.07 & 0 & 0.007 & & (1/4) & 0 & 0.024 & 0.123 & \\
N30d025M100 & - & 0.953 & 0.168 & 100 & & - & 0.155 & 0.934 & 100 \\ 
(0/5) & - & 0.042 & 0.063 & & (0/5) & - & 0.022 & 0.092 & \\ \hline
N50d015M100 & - & 1 & 0.129 & 93.33 & & - & 0.187 & 0.905 & 100 \\
(0/5) & - & 0 & 0.089 & & (0/5) & - & 0.099 & 0.095 & \\
N50d025M100 & - & 1 & 0.592 & 93.33 & & - & 0.448 & 1.312 & 100 \\
(0/5) & - & 0 & 0.147 & & (0/5) & - & 0.080 & 0.185 & \\ 
\bottomrule
\end{tabular}
}
\end{sc}
\end{small}
\end{center}
\vskip -0.1in
\end{table}


\begin{table}[t]
\caption{Computational results for problem instances with maximum edge weight $M_w = 50$, $K=2$ clusters, and at least $\sigma = 0.7$ vertices must be in at least one cluster.  
}
\label{tableK=2}
\vskip 0.15in
\begin{center}
\begin{small}
\begin{sc}
\resizebox{\columnwidth}{!}{
\begin{tabular}{lcccc|ccccc}
\toprule 
 &    \multicolumn{4}{c}{Min Cut} &     \multicolumn{5}{c}{Max Association}  \\
Data set 	& Opt & Gap  & r & Con & & Opt  & Gap & r & Con \\
\midrule
N15d015M50 & 0.196 & 0 & 0.061 & 100 && 0.203 & 0 & 0.246 & 100 \\
(5/0) & 0.067 & 0 & 0.048 && (5/0) & 0.058 & 0 & 0.041 & \\
N15d025M50 & 0.180 & 0 & 0.014 & 90.00& & 0.256 & 0 & 0.264 & 100 \\
(5/0) & 0.038 & 0 & 0.027 & &(5/0) & 0.066 & 0 & 0.079 & \\
N15d05M50 & 2.218 & 0 & 0.090 & 100 && 1.660 & 0 & 0.313 & 100 \\
(5/0) & 1.594 & 0 & 0.032 && (5/0) & 0.670 & 0 & 0.056 & \\ \hline
N20d015M50 & 0.211 & 0 & 0.014 & 90.00 && 0.699 & 0 & 0.214 & 100 \\
(5/0) & 0.067 & 0 & 0.032 && (5/0) & 0.288 & 0 & 0.051 & \\
N20d025M50 & 1.181 & 0 & 0.011 & 100 && 2.537 & 0 & 0.283 & 100 \\
(5/0) & 0.680 & 0 & 0.015 & &(5/0) & 1.288 & 0 & 0.064 & \\
N20d05M50 & 46.63 & 0 & 0.122 & 100& & 61.40 & 0 & 0.352 & 100 \\
(5/0) & 29.13 & 0 & 0.064 && (5/0) & 26.19 & 0 & 0.037 & \\ \hline
N30d015M50 & 1.548 & 0 & 0.000 & 80.00 && 12.10 & 0 & 0.248 & 100 \\
(5/0) & 0.461 & 0 & 0.000 && (5/0) & 4.455 & 0 & 0.051 & \\
N30d025M50 & 63.92 & 0 & 0.023 & 100 && 169.4 & 0.006 & 0.323 & 100 \\
(5/0) & 71.98 & 0 & 0.015 && (4/1) & 185.0 & 0 & 0.050 & \\ \hline
N50d015M50 & 0 & 0.887 & 0.023 & 100& & 0 & 0.082 & 0.287 & 100 \\
(0/5) & 0 & 0.162 & 0.015 & &(0/5) & 0 & 0.014 & 0.032 & \\
N50d025M50 & 0 & 0.997 & 0.111 & 100 && 0 & 0.177 & 0.354 & 100 \\
(0/5) & 0 & 0.002 & 0.028 && (0/5) & 0 & 0.014 & 0.055 & \\
\bottomrule
\end{tabular}
}
\end{sc}
\end{small}
\end{center}
\vskip -0.1in
\end{table}


\begin{table}[t]
\caption{Computational results for problem instances with maximum edge weight $M_w = 50$, $K=4$ clusters, and at least $\sigma = 0.7$ vertices must be in at least one cluster.  
}
\label{tableK=4}
\vskip 0.15in
\begin{center}
\begin{small}
\begin{sc}
\resizebox{\columnwidth}{!}{
\begin{tabular}{lcccc|ccccc}
\toprule 
 &    \multicolumn{4}{c}{Min Cut} &   \multicolumn{5}{c}{Max Association}  \\
Data set 	& Opt & Gap  & r & Con & & Opt  & Gap & r & Con \\
\midrule
N15d015M50 & 3.453 & 0 & 0.566 & 75.00 && 92.08 & 0 & 1.614 & 65.00 \\
(5/0) & 2.826 & 0 & 0.343 && (5/0) & 68.72 & 0 & 0.260 & \\
N15d025M50 & 75.79 & 0 & 0.252 & 75.00 && 300.5 & 0.053 & 1.553 & 75.00 \\
(5/0) & 100.2 & 0 & 0.141 & &(2/3) & 271.8 & 0.049 & 0.358 & \\
N15d05M50 & 0 & 0.617 & 0.863 & 75.00 && 0 & 0.232 & 1.737 & 75.00 \\
(0/5) & 0 & 0.090 & 0.294 & &(0/5) & 0 & 0.015 & 0.236 & \\ \hline
N20d015M50 & 83.63 & 0 & 0.104 & 75.00 && 41.54 & 0.035 & 1.287 & 65.00 \\
(5/0) & 128.8 & 0 & 0.105 && (1/4) & 0 & 0.004 & 0.319 & \\
N20d025M50 & 0 & 0.724 & 0.322 & 75.00 && 0 & 0.136 & 1.821 & 75.00 \\
(0/5) & 0 & 0.252 & 0.169 & &(0/5) & 0 & 0.020 & 0.143 & \\
N20d05M50 & 0 & 0.988 & 1.112 & 75.00 && 0 & 0.346 & 2.201 & 75.00 \\
(0/5) & 0 & 0.007 & 0.446 &&(0/5) & 0 & 0.057 & 0.306 & \\ \hline
N30d015M50 & 237.4 & 0.943 & 0.060 & 75.00 && 0 & 0.080 & 1.670 & 75.00 \\
(1/4) & 0 & 0.085 & 0.070 && (0/5) & 0 & 0.029 & 0.180 & \\
N30d025M50 & 0 & 1 & 0.365 & 75.00& & 0 & 0.248 & 1.988 & 75.00 \\
(0/5) & 0 & 0 & 0.252 && (0/5) & 0 & 0.034 & 0.267 & \\ \hline
N50d015M50 & 0 & 1 & 0.794 & 65.00 && 0 & 0.447 & 1.955 & 70.00 \\
(0/5) & 0 & 0 & 0.468 && (0/5) & 0 & 0.102 & 0.162 & \\
\bottomrule
\end{tabular}
}
\end{sc}
\end{small}
\end{center}
\vskip -0.1in
\end{table}


\begin{table}[t]
\caption{Computational results for problem instances with maximum edge weight $M_w = 50$, $K=3$ clusters, and tested $\sigma = 0.5$ versus  $\sigma = 0.8$.  
}
\label{tableSigmas}
\vskip 0.15in
\begin{center}
\begin{small}
\begin{sc}
\resizebox{\columnwidth}{!}{
\begin{tabular}{lcccc|ccccc}
\toprule 
 &    \multicolumn{4}{c}{Min Cut} &   \multicolumn{5}{c}{Max Association}  \\
Data set 	& Opt & Gap  & r & Con & & Opt  & Gap & r & Con \\
\midrule
N15d015M50 & 0.574 & 0 & 0.075 & 100 & & 0.698 & 0 & 0.363 & 100\\
(5/0) & 0.305 & 0 & 0.100 & & (5/0) & 0.206 & 0 & 0.334 & \\
N15d025M50 & 0.482 & 0 & 0.000 & 100 & & 1.403 & 0 & 0.140 & 100\\
(5/0) & 0.389 & 0 & 0.000 & &(5/0) & 0.159 & 0 & 0.096 & \\
N15d05M50 & 24.29 & 0 & 0.258 & 100 && 116.2 & 0 & 0.391 & 100\\
(5/0) & 12.43 & 0 & 0.174 && (5/0) & 33.96 & 0 & 0.116 &   \\ \hline
N20d015M50 & 1.019 & 0 & 0.000 & 100 & & 5.951 & 0 & 0.034 & 100\\
(5/0) & 0.737 & 0 & 0.000 & &  (5/0) & 8.141 & 0 & 0.025 &\\
N20d025M50 & 12.31 & 0 & 0.018 & 100 & & 149.4 & 0 & 0.237 & 100\\
(5/0) & 10.26 & 0 & 0.016 & &  (5/0) & 199.2 & 0 & 0.083 &  \\
N20d05M50 & 55.45 & 0.439 & 0.277 & 100 &   & 0 & 0.713 & 0.620 & 100\\
(1/4) & 0 & 0.168 & 0.228 & &  (0/5) & 0 & 0.080 & 0.281 &  \\ \hline
N30d015M50 & 3.175 & 0 & 0.000 & 93.33 &  & 357.0 & 0.813 & 0.086 & 93.33\\
(5/0) & 2.334 & 0 & 0.000 & &  (4/1) & 196.5 & 0 & 0.067 &\\
N30d025M50 & 47.12 & 0 & 0.001 & 100 & & 0 & 0.959 & 0.265 & 100\\
(5/0) & 53.50 & 0 & 0.002 & &(0/5) & 0 & 0.026 & 0.133 &  \\ \hline
N50d015M50 & 209.7 & 1 & 0.007 & 93.33 &  & 0 & 1 & 0.313 & 100 \\
(4/1) & 212.0 & 0 & 0.016 && (0/5) & 0 & 0 & 0.137 &\\
N50d025M50 & 0 & 1 & 0.219 & 100 &   & 0 & 1 & 0.903 & 93.33\\
(0/5) & 0 & 0 & 0.109 && (0/5) & 0 & 0 & 0.317 & \\
\bottomrule
\end{tabular}
}
\end{sc}
\end{small}
\end{center}
\vskip -0.1in
\end{table}


\section{Conclusions and future research directions}
\label{Sec:Conc}
We proposed and developed a polynomial-size MILP model for the SGCP. As far as we aware, this is the first approach that simultaneously allocates membership proportion (the $x_{ic}$ values) for vertices that lie in multiple clusters, and that enforces an equal balance of the clusters so that the sum of vertex memberships over all clusters are roughly the same. Compared to CFinder (\cite{Palla2005}, \cite{Derenyi2005}, 
\cite{Adamcsek2006}), the clusters found in our method are not limited to $k$-clique neighbourhoods. Compared to MaxMax (\cite{Hope2013}), our method can produce non-trivial clusters even for a connected unweighted graph. 
An obvious future research direction is to perform a thorough numerical experiment on all parameters used in the model. One can consider alternative formulations, e.g., a branch-and-cut approach, using constraints to cut off infeasible solutions (namely, solutions with disconnected clusters); or, a branch-and-price approach, using exponentially many variables each representing a feasible cluster (in which case connectivity is guaranteed). Even though the cut separation and column generation subproblems themselves are expected to be  NP-hard, heuristic approaches can be derived for speedy execution. Further, one may consider an adaptive approach to find the optimal number of clusters, instead of iteratively optimising over different values of $K$. 

\clearpage 
\bibliographystyle{named}
\bibliography{MILP-MY}

\begin{thebibliography}{}

\bibitem[\protect\citeauthoryear{Adamcsek \bgroup \em et al.\egroup
  }{2006}]{Adamcsek2006}
Balázs Adamcsek, Gergely Palla, Illés Farkas, Imre Derényi, and Tamás
  Vicsek.
\newblock Cfinder: Locating cliques and overlapping modules in biological
  networks.
\newblock {\em Bioinformatics (Oxford, England)}, 22:1021--3, 05 2006.

\bibitem[\protect\citeauthoryear{Bertsimas and Shioda}{2007}]{Bertsimas2007}
Dimitris Bertsimas and Romy Shioda.
\newblock Classification and regression via integer optimization.
\newblock {\em Oper. Res.}, 55(2):252--271, March 2007.

\bibitem[\protect\citeauthoryear{Biemann}{2006}]{Biemann2006}
Christian Biemann.
\newblock Chinese whispers - an efficient graph clustering algorithm and its
  application to natural language processing problems.
\newblock 2006.

\bibitem[\protect\citeauthoryear{Clauset \bgroup \em et al.\egroup
  }{2004}]{Clauset2004}
Aaron Clauset, M.~E.~J. Newman, and Cristopher Moore.
\newblock Finding community structure in very large networks.
\newblock {\em Phys. Rev. E}, 70:066111, Dec 2004.

\bibitem[\protect\citeauthoryear{Dai and Charkhgard}{2018}]{Dai2018}
Rui Dai and Hadi Charkhgard.
\newblock A two-stage approach for bi-objective integer linear programming.
\newblock {\em Oper. Res. Lett.}, 46:81--87, 2018.

\bibitem[\protect\citeauthoryear{Der\'enyi \bgroup \em et al.\egroup
  }{2005}]{Derenyi2005}
Imre Der\'enyi, Gergely Palla, and Tam\'as Vicsek.
\newblock Clique percolation in random networks.
\newblock {\em Phys. Rev. Lett.}, 94:160202, Apr 2005.

\bibitem[\protect\citeauthoryear{Gilpin \bgroup \em et al.\egroup
  }{2013}]{Gilpin2013}
Sean Gilpin, Siegried Nijssen, and Ian Davidson.
\newblock Formalizing hierarchical clustering as integer linear programming,
  2013.

\bibitem[\protect\citeauthoryear{Girvan and Newman}{2002}]{Girvan2002}
M.~Girvan and M.~E.~J. Newman.
\newblock Community structure in social and biological networks.
\newblock {\em Proceedings of the National Academy of Sciences},
  99(12):7821--7826, 2002.

\bibitem[\protect\citeauthoryear{Hope and Keller}{2013}]{Hope2013}
David Hope and Bill Keller.
\newblock Maxmax: A graph-based soft clustering algorithm applied to word sense
  induction.
\newblock In Alexander Gelbukh, editor, {\em Computational Linguistics and
  Intelligent Text Processing}, pages 368--381, Berlin, Heidelberg, 2013.
  Springer Berlin Heidelberg.

\bibitem[\protect\citeauthoryear{Liu and Foroushani}{2016}]{Liu2016}
Ying Liu and Amir Foroushani.
\newblock Pfc: An efficient soft graph clustering method for ppi networks based
  on purifying and filtering the coupling matrix.
\newblock In De-Shuang Huang, Vitoantonio Bevilacqua, and Prashan Premaratne,
  editors, {\em Intelligent Computing Theories and Application}, pages
  247--257, Cham, 2016. Springer International Publishing.

\bibitem[\protect\citeauthoryear{Marinka~Zitnik and
  Leskovec}{2018}]{biosnapnets}
Sagar~Maheshwari Marinka~Zitnik, Rok~Sosi\v{c} and Jure Leskovec.
\newblock {BioSNAP Datasets}: {Stanford} biomedical network dataset collection.
\newblock \url{http://snap.stanford.edu/biodata}, August 2018.

\bibitem[\protect\citeauthoryear{Miller \bgroup \em et al.\egroup
  }{1960}]{Miller1960}
C.~E. Miller, A.~W. Tucker, and R.~A. Zemlin.
\newblock Integer programming formulation of traveling salesman problems.
\newblock {\em Journal of the ACM}, 7(4):326--329, October 1960.

\bibitem[\protect\citeauthoryear{Miyauchi \bgroup \em et al.\egroup
  }{2018}]{Miyauchi2018}
Atsushi Miyauchi, Tomohiro Sonobe, and Noriyoshi Sukegawa.
\newblock Exact clustering via integer programming and maximum satisfiability,
  2018.

\bibitem[\protect\citeauthoryear{Palla \bgroup \em et al.\egroup
  }{2005}]{Palla2005}
Gergely Palla, Imre Der{\'{e}}nyi, Ill{\'{e}}s Farkas, and Tamás Vicsek.
\newblock Uncovering the overlapping community structure of complex networks in
  nature and society.
\newblock {\em Nature}, 435(7043):814--818, June 2005.

\bibitem[\protect\citeauthoryear{Pinney and Westhead}{2006}]{Pinney2006}
John~W. Pinney and David~R. Westhead.
\newblock Betweenness-based decomposition methods for social and biological
  networks.
\newblock 2006.

\bibitem[\protect\citeauthoryear{Rosvall and Bergstrom}{2008}]{Rosvall2008}
Martin Rosvall and Carl~T. Bergstrom.
\newblock Maps of random walks on complex networks reveal community structure.
\newblock {\em Proceedings of the National Academy of Sciences},
  105(4):1118--1123, 2008.

\bibitem[\protect\citeauthoryear{Sağlam \bgroup \em et al.\egroup
  }{2006}]{Saglam2006}
Burcu Sağlam, F.~Sibel Salman, Serpil Sayın, and Metin Türkay.
\newblock A mixed-integer programming approach to the clustering problem with
  an application in customer segmentation.
\newblock {\em European Journal of Operational Research}, 173(3):866 -- 879,
  2006.

\bibitem[\protect\citeauthoryear{Schaeffer}{2007}]{Schaeffer2007}
Satu~Elisa Schaeffer.
\newblock Graph clustering.
\newblock {\em Computer Science Review}, 1(1):27 -- 64, 2007.

\bibitem[\protect\citeauthoryear{sna}{}]{snapnets}


\bibitem[\protect\citeauthoryear{Support}{2018}]{Cplex}
IBM Support.
\newblock Cplex optimization studio v12.8.
\newblock \url{http://www-01.ibm.com/support/docview.wss?uid=swg27050618},
  2018.

\bibitem[\protect\citeauthoryear{Ustalov \bgroup \em et al.\egroup
  }{2018}]{Ustalov2018}
Dmitry Ustalov, Alexander Panchenko, Chris Biemann, and Simone~Paolo Ponzetto.
\newblock Local-global graph clustering with applications in sense and frame
  induction.
\newblock {\em CoRR}, abs/1808.06696, 2018.

\bibitem[\protect\citeauthoryear{Ye}{2007}]{Ye2007}
M.~Ye.
\newblock An integer programming clustering approach with application to
  recommendation systems.
\newblock Master's thesis, Iowa State University, 2007.

\end{thebibliography}

\end{document}